\documentclass[conference]{IEEEtran}
\IEEEoverridecommandlockouts
\usepackage{amsmath}
\usepackage{amssymb}
\usepackage{amsfonts}
\usepackage{algorithm}
\usepackage{algpseudocode}
\usepackage{comment}
\usepackage{dsfont}
\usepackage{float}
\usepackage{cite}
\usepackage{graphicx}
\usepackage[left=0.625in,right=0.625in,top=0.7in,bottom=1in]{geometry}
\usepackage{epsfig}
\usepackage{subfigure}
\usepackage{psfrag}
\usepackage{xcolor}
\usepackage{url}
\allowdisplaybreaks
\usepackage[colorlinks,linkcolor=black,urlcolor=black,anchorcolor=black,citecolor=black,hyperfootnotes=true]{hyperref}

\begin{document}

\title{Trilateration-Based Device-Free Sensing: Two Base Stations and One Passive IRS Are Sufficient}	
\author{%
  \IEEEauthorblockN{Qipeng Wang, Liang Liu, Shuowen Zhang, and Francis C.M. Lau}
  \IEEEauthorblockA{Department of Electronic and Information Engineering, The Hong Kong Polytechnic University\\
                    E-mails: qipeng.wang@connect.polyu.hk, $\{$liang-eie.liu,shuowen.zhang,francis-cm.lau$\}$@polyu.edu.hk}

}

\maketitle	

\begin{abstract}
The classic trilateration technique can localize each target based on its distances to three anchors with known coordinates. Usually, this technique requires all the anchors and targets, e.g., the satellites and the mobile phones in Global Navigation Satellite System (GNSS), to actively transmit/receive radio signals such that the delay of the one-way radio signal propagated between each anchor and each target can be measured. Excitingly, this paper will show that the trilateration technique can be generalized to the scenario where one of the three anchors and all the targets merely reflect the radio signals passively as in radar networks, even if the propagation delay between the passive IRS and the passive targets is difficult to be measured directly, and the data association issue for multi-sensor multi-target tracking arises. Specifically, we consider device-free sensing in a cellular network consisting of two base stations (BSs), one passive intelligent reflecting surface (IRS), and multiple passive targets, to realize integrated sensing and communication (ISAC). The two BSs transmit the orthogonal frequency division multiplexing (OFDM) signals in the downlink and estimate the locations of the targets based on their reflected signals via/not via the IRS. We propose an efficient trilateration-based strategy that can first estimate the distances of each target to the two BSs and the IRS and then localize the targets. Numerical results show that the considered networked sensing architecture with heterogenous anchors can outperform its counterpart with three BSs.
\end{abstract}

\section{Introduction}\label{Sec:intro}
Recently, there is a trend in both academia and industry to achieve integrated sensing and communication (ISAC) in the future 6G cellular network via utilizing a common radio spectrum and the same hardware platform \cite{liu2020joint,zheng2019radar}. As a result, when a new 6G technology appears, we should not only evaluate its effectiveness in communication, but also understand its potential role in sensing. For communication, recently, there has been a flurry of research activities in using intelligent reflecting surface (IRS) to enhance network throughput in the 6G era \cite{Wu21}. One interesting question thus arises: is IRS also beneficial for sensing? The main contribution of this paper is to provide an affirm answer to the above question.

Compared to the large body of research in IRS-assisted communication, the investigation of IRS-assisted sensing is still in its infancy. Along this line, the basic idea is to view each IRS as an anchor with known location and localize the targets based on their relative position to the IRSs and the base stations (BSs). Specifically, under the device-based sensing setup, where the targets can transmit/receive radio signals actively such that their location can be estimated based on the one-way radio signals to/from the anchors, the IRS-assisted sensing has been studied for the time-of-arrival (TOA) approach, the angle-of-arrival (AOA) or angle-of-departure (AOD) approach, as well as the received-signal-strength (RSS) approach \cite{Dardari22,Zhang21,Elzanaty21,Lin22,Song21}. However, in most ISAC applications, we need to perform device-free sensing, where the targets just reflect the signals passively to the anchors such that we have to localize them based on the round-trip signals. Recently, \cite{dvc} showed that the BSs in the cellular network can collaboratively perform networked sensing to localize the passive targets based on the trilateration method, where each target is localized via its distances to multiple BSs. Compared to device-based sensing, the main challenge here lies in data association \cite{Mahler07}: since all the targets reflect the same signals to the BSs, it is non-trivial to match each estimated distance to the right target. Interestingly, it was shown in \cite{dvc} that when all the anchors are active BSs, the data association issue can be effectively resolved. However, how to perform trilateration-based device-free sensing if some BSs are replaced by the passive IRSs is still an open problem in the literature.

In this paper, we consider device-free sensing in a cellular network consisting of two BSs, one IRS, and several passive targets close to the IRS, as shown in Fig. \ref{fig1}. In the downlink, the two BSs transmit the orthogonal frequency division multiplexing (OFDM) signals to the mobile users, which will be reflected by the targets back to the BSs via/not via the IRS. A two-phase protocol is utilized to localize the targets based on their reflected signals. In the first phase, we apply the OFDM channel estimation technique to obtain the delay (thus the range) of each BS-target link and BS-IRS-target-BS link; while in the second phase, the data association and the target location are estimated based on the above range information. The key observation of the proposed scheme is that the distance between a target and the IRS can be obtained from either the link from BS 1 to the IRS to the target to BS 1 or that from BS 2 to the IRS to the target to BS 2. As a result, if some matching solution results in significant difference in estimating the distance between some target and the IRS based on the above two approaches, this solution cannot be the true data association solution. Such a property enables a powerful data association algorithm in this paper. In practice, as compared to the BSs, the low-cost IRSs can be deployed with a much higher density such that each target can be sensed by covered by multiple anchors. As a result, our theoretical results show that networked sensing with both active anchors (BSs) and passive anchors (IRSs) is a promising solution in future 6G network.
	
\section{System Model}\label{Sec:SysMod}
\begin{figure}[t]
	\centering
	\includegraphics[width=8.5cm]{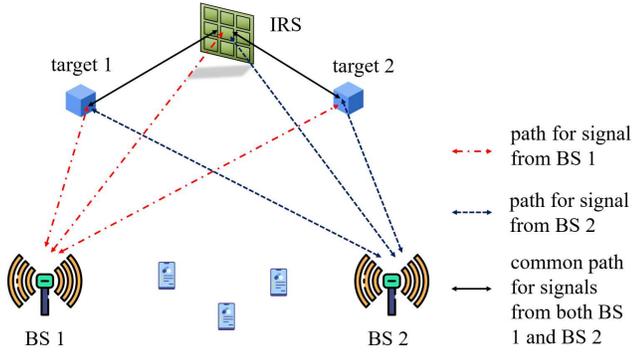}
	\vspace{-3mm}
	\caption{System model of an IRS-assisted ISAC cellular network.}\label{fig1}
	\vspace{-7mm}
\end{figure}

We consider an IRS-assisted OFDM-based ISAC system as illustrated in Fig. \ref{fig1}, which consists of two BSs, one IRS equipped with $I$ passive elements, $U$ mobile communication users to be served, and $K$ targets to be localized. The number of OFDM sub-carriers is denoted by $N$, and the sub-carrier spacing is denoted by $\Delta f$ Hz, thus the overall channel bandwidth is $B=N\Delta f$ Hz. Moreover, let $\mathcal{N}_1$ and $\mathcal{N}_2$ denote the sets of sub-carriers allocated to BS 1 and BS 2, respectively, where $\mathcal{N}_1\bigcap \mathcal{N}_2=\emptyset$ and $\mathcal{N}_1\bigcup\mathcal{N}_2=\{1,\dots,N\}$.

Since IRS-assisted communication has been widely studied in the literature, this paper focuses on IRS-assisted sensing to realize ISAC in the considered OFDM-based system. Let $(x_m^{\text{A}}, y_m^{\text{A}})$ and $(x^{\text{I}}, y^{\text{I}})$ in meter denote the locations of the $m$-th BS ($m=1,2$) and the IRS in the two-dimensional (2D) Cartesian coordinate system, respectively, which are fixed and known. Therefore, \emph{BS 1, BS 2, as well as the IRS can serve as three anchors} for sensing the environment. Moreover, let $(x_k^{\text{T}}, y_k^{\text{T}})$ in meter denote the location of the $k$-th target, $k=1,\ldots,K$. Define
\begin{align}
 d_{m,k}^{\text{AT}}=&\sqrt{(x_k^{\text{T}}-x_m^{\text{A}})^2+(y_k^{\text{T}}-y_m^{\text{A}})^2}, \label{eq:range AT} \\
 d_{m}^{\text{AI}}=&\sqrt{(x^{\text{I}}-x_m^{\text{A}})^2+(y^{\text{I}}-y_m^{\text{A}})^2}, \label{eq:range AI} \\
 d_k^{\text{IT}}=&\sqrt{(x_k^{\text{T}}-x^{\text{I}})^2+(y_k^{\text{T}}-y^{\text{I}})^2}, ~ \forall m, k, \label{eq:range IT}
\end{align} in meter as the distance between BS $m$ and target $k$, the distance between BS $m$ and the IRS, and the distance between the IRS and target $k$, respectively.

To estimate the target locations $(x_k^{\text{T}}, y_k^{\text{T}})$'s, the two BSs first send out downlink OFDM communication signals, and then collaboratively localize the targets based on their reflected signals as well as the IRS's reflected signals. Specifically, we let $Q$ denote the number of OFDM symbols in one resource block (RB), e.g., $Q=7$ in 4G LTE systems, and focus our study on the transmission in one RB. In each $q$-th OFDM symbol, let $s_{m,n}^{(q)}$ denote the information symbol sent from the $m$-th BS at the $n$-th sub-carrier, with $\mathbb{E}[|s_{m,n}^{(q)}|^2]=1$ if $n\in \mathcal{N}_m$ and $s_{m,n}^{(q)}=0$ if $n\notin \mathcal{N}_m$. We further let $\boldsymbol{s}_m^{(q)}=[s_{m,1}^{(q)},\ldots,s_{m,N}^{(q)}]^T$ denote the information symbols from the $m$-th BS over all the $N$ sub-carriers in the $q$-th OFDM symbol. Let $p_m$ denote the common transmit power at BS $m$ at all sub-carriers. Thus, the time-domain downlink OFDM signal transmitted by the $m$-th BS in the $q$-th OFDM symbol is given by
\begin{equation}\label{eq:sg_tm}
    \boldsymbol{\chi}_m^{(q)}=[\chi_{m,1}^{(q)},\ldots,\chi_{m,N}^{(q)}]^T
    =\boldsymbol{W}^H \sqrt{p_m}\boldsymbol{s}_{m}^{(q)}, ~ \forall m,q,
\end{equation}where $\chi_{m,n}^{(q)}$ denotes the $n$-th sample transmitted by the $m$-th BS in the $q$-th OFDM symbol, and $\boldsymbol{W}\in \mathbb{C}^{N\times N}$ denotes the $N\times N$ discrete Fourier transform (DFT) matrix.

Similar to the works focusing on the classic trilateration-based localization theory \cite{Torrieri84,Mao07,dvc}, we assume that there only exists line-of-sight (LoS) links in the BS-target channels, BS-IRS channels, and IRS-target channels. Thus, each BS $m$ receives the reflected signals at its allocated sub-carriers $\mathcal{N}_m$ via three types of links: the links from BS $m$ via targets to BS $m$, the link from BS $m$ via the IRS to BS $m$, and the links from BS $m$ via the IRS and targets to BS $m$.\footnote{The other links involving more reflections, e.g., the links from BS $m$ via the IRS, targets, and the IRS again to BS $m$, are ignored since they are too weak as compared to the above links.} Note that these links automatically form a multi-path environment. Specifically, define
\begin{align}
\boldsymbol{h}^{{\rm ATA}}_m=&[h^{{\rm ATA}}_{m,1},\ldots,h^{{\rm ATA}}_{m,L}]^T, \label{eq:ATA} \\
\boldsymbol{h}^{{\rm AIA}}_{m,i}=&[h^{{\rm AIA}}_{m,i,1},\ldots,h^{{\rm AIA}}_{m,i,L}]^T, \label{eq:AIA} \\
\boldsymbol{h}^{{\rm AITA}}_{m,i}=&[h^{{\rm AITA}}_{m,i,1},\ldots,h^{{\rm AITA}}_{m,i,L}]^T, ~ \forall m,i, \label{eq:AITA}
\end{align}
as the $L$-tap baseband equivalent multi-path channels for the links from BS $m$ via targets to BS $m$, the cascaded links from BS $m$ via IRS reflecting element $i$ to BS $m$, and the cascaded links from BS $m$ via IRS element $i$ and targets to BS $m$, respectively, where $L$ denotes the maximum number of paths determined by the delay spread. Note that $h^{{\rm ATA}}_{m,l}\neq 0$ holds if and only if there exists a target $\bar{k}_{m,l}$ such that the delay of the propagation path from BS $m$ via target $\bar{k}_{m,l}$ to BS $m$ is within $l-1$ OFDM samples and $l$ OFDM samples, i.e.,
\begin{align}\label{eq:dis 1}
\frac{(l-1)c_0}{2N\Delta f}\leq d_{m,\bar{k}_{m,l}}^{\rm AT}\leq \frac{lc_0}{2N \Delta f},
\end{align}
where $c_0$ denotes the speed of light in meter/second. Moreover, $h^{{\rm AIA}}_{m,i,l}\neq 0$, $\forall i$, holds if and only if the distance between BS $m$ and the IRS satisfies
\begin{align}\label{eq:dis 3}
\frac{(l-1)c_0}{2N\Delta f}\leq d_{m}^{\text{AI}}\leq \frac{lc_0}{2N \Delta f}.
\end{align}
Lastly, $h^{{\rm AITA}}_{m,i,l}\neq 0$, $\forall i$, holds if and only if there exists a target $\hat{k}_{m,l}$ satisfying
\begin{align}\label{eq:dis 2}
\frac{(l-1)c_0}{N\Delta f}\leq d_{m,\hat{k}_{m,l}}^{{\rm AITA}} \leq \frac{lc_0}{N \Delta f},
\end{align}where
\begin{align}\label{eq:dis}
d_{m,\hat{k}_{m,l}}^{{\rm AITA}}=d_m^{{\rm AI}}+d_{\hat{k}_{m,l}}^{{\rm IT}}+d_{m,\hat{k}_{m,l}}^{\rm AT},
\end{align}
in meter denotes the range (length) of the path from BS $m$ via the IRS and target $\hat{k}_{m,l}$ to BS $m$.

Moreover, denote $\phi_i^{(q)}\in \mathbb{C}$ as the reflection coefficient of the $i$-th IRS reflecting element in the $q$-th OFDM symbol duration, with $|\phi_i^{(q)}|=1,\forall i,q$, due to the passiveness of the IRS. The overall $L$-tap multi-path channel of BS $m$ for the $q$-th OFDM symbol contributed by all the targets and the IRS can be expressed as
\begin{align}\label{eq:channel}
	\boldsymbol{h}_m^{(q)}&=[h_{m,1}^{(q)},\ldots,h_{m,L}^{(q)}]^T \nonumber \\ & =\boldsymbol{h}^{{\rm ATA}}_m+\sum_{i=1}^I\phi_i^{(q)}(\boldsymbol{h}^{{\rm AIA}}_{m,i}+\boldsymbol{h}^{{\rm AITA}}_{m,i}), ~ \forall m, q.
\end{align}
Thus, in the $q$-th OFDM symbol duration, the signals received by BS $m$ at its allocated sub-carriers $\mathcal{N}_m$ is given by
\begin{align}\label{eq:rcvd_sgnl}
    \boldsymbol{y}_m^{(q)}&=[y_{m,\mathcal{N}_{m}(1)}^{(q)},\ldots,y_{m,\mathcal{N}_{m}(|\mathcal{N}_{m}|)}^{(q)}]^T\notag\\
    &=\!\sqrt{p_m}\text{diag}(\tilde{\boldsymbol{s}}_m^{(q)})\boldsymbol{E}_m\boldsymbol{h}_m^{(q)}\!+\!\boldsymbol{z}_m^{(q)},~\forall m,q,
\end{align}
where $\tilde{\boldsymbol{s}}_m^{(q)}=[s_{m,\mathcal{N}_m(1)}^{(q)},\ldots,s_{m,\mathcal{N}_m(|\mathcal{N}_m|)}^{(q)}]^T\in \mathbb{C}^{|\mathcal{N}_m|\times 1}$ is the collection of information symbols sent by BS $m$ at all its allocated sub-carriers in the $q$-th OFDM symbol, $\boldsymbol{E}_m\in \mathbb{C}^{|\mathcal{N}_m|\times L}$ with the $(n,l)$-th element given by $E_{m,n,l}=e^{-\frac{j2\pi(\mathcal{N}_m(n)-1)(l-1)}{N}}$, and $\boldsymbol{z}_m^{(q)}\sim \mathcal{CN}(\boldsymbol{0},\sigma^2\boldsymbol{I})$ denotes the receiver noise at BS $m$ in the $q$-th OFDM symbol duration.
\section{Two-Phase Localization Protocol}\label{sec:Two-Phase Localization Protocol}
In this paper, we propose a two-phase protocol to localize the $K$ targets based on the received signals given in (\ref{eq:rcvd_sgnl}). In the first phase, the delay (thus the range) of each BS-target-BS link and each BS-IRS-target-BS link will be estimated based on the signals received by the two BSs; while in the second phase, the location of each target can be estimated based on the above range information using the trilateration method.

Specifically, range estimation in the first phase (Phase I) is performed based on the following philosophy. Note that $h_{m,l}^{{\rm ATA}}\neq 0$ (or $h_{m,i,l}^{{\rm AITA}}\neq 0$, $\forall i$) holds if and only if (\ref{eq:dis 1}) (or (\ref{eq:dis 2})) is true for some target $\bar{k}_{m,l}$ (or $\tilde{k}_{m,l}$). In a broadband communication system with a very large bandwidth $B=N\Delta f$, the left-hand side and the right-hand side of (\ref{eq:dis 1}) or (\ref{eq:dis 2}) are approximately the same. Therefore, we can first estimate the non-zero coefficients of the $L$-tap multi-path channels based on the received signal (\ref{eq:rcvd_sgnl}), and then estimate the range of an BS-target path and that of an BS-IRS-target-BS path accurately based on (\ref{eq:dis 1}) and (\ref{eq:dis 2}), respectively.

Next, in the second phase (Phase II), we localize the targets based on the trilateration method. According to (\ref{eq:dis}), for a target $k$, its distance to the IRS can be obtained from
\begin{align}\label{eq:dis 3}
d_k^{\text{IT}}\!=\!d_{1,k}^{{\rm AITA}}-d_1^{{\rm AI}}-d_{1,k}^{\rm AT}\!=\!d_{2,k}^{{\rm AITA}}-d_2^{{\rm AI}}-d_{2,k}^{\rm AT}, ~ \forall k.
\end{align}Note that $d_1^{{\rm AI}}$ and $d_2^{{\rm AI}}$ are known because the locations of the BSs and the IRS are known, while $d_{1,k}^{\rm AT}$ and $d_{2,k}^{\rm AT}$ can be estimated in Phase I. As a result, the distance between the IRS and each target can be obtained based on (\ref{eq:dis 3}). Theoretically speaking, the location of each target $k$ can be estimated based on the trilateration method with the knowledge about its distances to BS $1$, BS $2$, and the IRS. However, in practice, the main challenge for the above localization approach lies in the data association issue arising from device-free sensing \cite{dvc}. Specifically, as shown in (\ref{eq:rcvd_sgnl}), at the sub-carriers allocated to each BS $m$, all the targets reflect the same signal, i.e., the signal sent by BS $m$, back to this BS. Consequently, if there exists an $l$ such that $h^{{\rm ATA}}_{m,l}\neq 0$ or $h^{{\rm AITA}}_{m,i,l}\neq 0$, we do not directly know which target contributes to this BS-target-BS or BS-IRS-target-BS link, i.e., $\bar{k}_{m,l}$ in (\ref{eq:dis 1}) and $\hat{k}_{m,l}$ in (\ref{eq:dis 2}) are unknown. This indicates that although each BS can obtain rich range information in Phase I based on (\ref{eq:dis 1}) and (\ref{eq:dis 2}), it does not know how to match each range to the right target for localization. Therefore, in Phase II, we have to first perform data association to estimate $\bar{k}_{m,l}$'s in (\ref{eq:dis 1}) and $\hat{k}_{m,l}$'s in (\ref{eq:dis 2}), and then localize each target based on its matched distances to BS $1$, BS $2$, and the IRS.

It is worth noting that in our paper, the key for the data association algorithm design is (\ref{eq:dis 3}): for the right data association solution, (\ref{eq:dis 3}) should hold for all the targets. Here, the IRS, whose reflected signals contribute to the received signals of both of the two BSs at their assigned sub-carriers as shown in (\ref{eq:rcvd_sgnl}), is the core anchor for determining the data association solution. This is because different from the active anchors, i.e., the two BSs, the distance from the IRS to any target $k$ can be calculated based on two approaches: $d_k^{{\rm IT}}=d_{1,k}^{{\rm AITA}}-d_1^{{\rm AI}}-d_{1,k}^{\rm AT}$ via the path involving BS $1$, or $d_k^{{\rm IT}}=d_{2,k}^{{\rm AITA}}-d_2^{{\rm AI}}-d_{2,k}^{\rm AT}$ via the path involving BS $2$. Therefore, given any data association solution, if there exists at least a target whose estimated distance to the IRS obtained from the path involving BS $1$ is quite different from that obtained from the path involving BS $2$, then this data association solution is wrong. This property enables a low-complexity and accurate data association algorithm as will be shown later in the paper.

In the following, we introduce each phase of the proposed protocol in more details.

\section{Phase I: Range Estimation}
In this section, we introduce the range estimation method in Phase I under the two-phase localization protocol. As discussed in Section \ref{sec:Two-Phase Localization Protocol}, in this phase, we should first estimate the non-zero components in the $L$-tap multi-path channels $\boldsymbol{h}_m^{(q)}$'s. From the compatibility perspective, the sensing function should be achieved in cellular networks without changing the communication protocols too much. For IRS-assisted communication, the on/off protocol is widely advocated in the literature \cite{Wu21,Wang20}. Under this protocol, in each OFDM RB, the IRS is off in the first OFDM symbol period such that the direct channels between the BSs and the mobile users can be estimated, while the IRS is on in the remaining OFDM symbol period such that the IRS-related channels can be estimated and the data can be decoded based on the estimated channels. In the following, we introduce how to estimate the non-zero components in $\boldsymbol{h}_m^{(q)}$'s for sensing based on the above on/off protocol.

Specifically, the IRS is off in the first OFDM symbol period of a RB, i.e., $\phi_i^{(1)}=0$, $\forall i$. According to (\ref{eq:rcvd_sgnl}), the signal received at BS $m$ at all its assigned sub-carriers of the first OFDM symbol is given as
\begin{align}\label{eq:first symbol}
\boldsymbol{y}_m^{(1)}=\sqrt{p_m}\text{diag}(\tilde{\boldsymbol{s}}_m^{(1)})\boldsymbol{E}_m\boldsymbol{h}^{{\rm ATA}}_m+\boldsymbol{z}_m^{(1)}, ~~~ m=1,2,
\end{align}which is merely contributed by the BS-target-BS links. Note that many coefficients in $\boldsymbol{h}^{{\rm ATA}}_m$'s are zero. As a result, we can estimate each $\boldsymbol{h}^{{\rm ATA}}_m$ based on the LASSO technique by solving the following problem
\begin{align} \hspace{-8pt} \mathop{\mathrm{min}}_{\boldsymbol{h}^{{\rm ATA}}_m}    0.5\left\|\!\boldsymbol{y}_m^{(1)}\!-\!\sqrt{p_m}\text{diag}(\tilde{\boldsymbol{s}}_m^{(1)})\boldsymbol{E}_m\boldsymbol{h}^{{\rm ATA}}_m\!\right\|_{{\rm F}}^2\!+\!\rho\!\left\|\!\boldsymbol{h}^{{\rm ATA}}_m\!\right\|_1, \label{eqn:problem 1}
\end{align}where $\rho\geq 0$ is a given coefficient.

Let $\bar{\boldsymbol{h}}_m^{{\rm ATA}}=[\bar{h}_{m,1}^{{\rm ATA}},\ldots,\bar{h}_{m,L}^{{\rm ATA}}]^T$ denote the optimal solution to the above convex problem, $m=1,2$. Based on $\bar{\boldsymbol{h}}_m^{{\rm ATA}}$'s, we need to estimate the support of each $\boldsymbol{h}^{{\rm ATA}}_m$ so as to obtain the range of each BS-target link as shown in (\ref{eq:dis 1}). In this paper, we adopt a threshold-based strategy to achieve the above goal. Specifically, with some given threshold $\delta_1>0$, define $\Phi_m^{{\rm I}}=\{l|\|\bar{h}_{m,l}^{{\rm ATA}}\|_2\geq \delta_1\}$, $m=1,2$. Then, for any $l\in \Phi_m^{{\rm I}}$, we declare that $h_{m,l}^{{\rm ATA}}\neq 0$ and there exists a target $\bar{k}_{m,l}$ whose distance to BS $m$, i.e., $d_{m,\bar{k}_{m,l}}^{\rm AT}$, satisfies (\ref{eq:dis 1}). In this case, we estimate $d_{m,\bar{k}_{m,l}}^{\rm AT}$ as the middle point of the range shown in (\ref{eq:dis 1}), i.e.,
\begin{align}\label{eq:range 1}
\bar{d}_{m,\bar{k}_{m,l}}^{\rm AT}=\frac{(l-1)c_0}{2N\Delta f}+\frac{c_0}{4N\Delta f}, ~ {\rm if} ~ l\in \Phi_m^{{\rm I}}.
\end{align}To summarize, after the ``off'' state of Phase I, each BS $m$ will have a set consisting of the estimated distances between the targets and it, which is denoted by
\begin{align}\label{eq:set 1}
\mathcal{D}_m^{{\rm AT}}=\{\bar{d}_{m,\bar{k}_{m,l}}^{\rm AT}|\forall l\in \Phi_m^{{\rm I}}\}, ~ m=1,2.
\end{align}

In the remaining OFDM symbol period of a RB, the IRS is on. Define $\boldsymbol{\bar{y}}_m^{(q)}=[\text{diag}(\tilde{\boldsymbol{s}}_m^{(q)})]^{-1}\boldsymbol{\bar{y}}_m^{(q)}$ and $\boldsymbol{\bar{z}}_m^{(q)}=[\text{diag}(\tilde{\boldsymbol{s}}_m^{(q)})]^{-1}\boldsymbol{\bar{z}}_m^{(q)}$, $m=1,2$ and $q=2,\ldots,Q$, as the effective received signal and noise at BS $m$ for the $q$-th OFDM symbol, respectively, where the effective received signal and noise at each assigned sub-carrier is normalized by the signal transmitted at this sub-carrier. According to (\ref{eq:rcvd_sgnl}), the collection of the effective received signals at BS $m$ for the remaining $Q-1$ OFDM symbols is given as
\begin{align}\label{eq:remaining symbol}
\boldsymbol{\bar{Y}}_m^{(2:Q)}&=[\boldsymbol{\bar{y}}_m^{(2)},\ldots,\boldsymbol{\bar{y}}_m^{(Q)}]^T\nonumber \\ &=\sqrt{p_m}\boldsymbol{E}_m\boldsymbol{H}^{(2:Q)}_m+\boldsymbol{\bar{Z}}_m^{(2:Q)}, ~~~ m=1,2,
\end{align}where $\boldsymbol{H}^{(2:Q)}_m=[\boldsymbol{h}_m^{(2)},\ldots,\boldsymbol{h}_m^{(Q)}]$ and $\boldsymbol{\bar{Z}}_m^{(2:Q)}=[\boldsymbol{\bar{z}}_m^{(2)},\ldots,\boldsymbol{\bar{z}}_m^{(Q)}]$.

Note that both $\boldsymbol{H}^{(2:Q)}_1$ and $\boldsymbol{H}^{(2:Q)}_2$ are row-sparse matrices, i.e., $\boldsymbol{g}_{m,l}=\boldsymbol{0}$ for many $l$, where $\boldsymbol{g}_{m,l}=[h_{m,l}^{(2)},\ldots,h_{m,l}^{(Q)}]^T$ denotes the $l$-th row of $\boldsymbol{H}^{(2:Q)}_m$. This is because if there is no path causing a delay of $l$ OFDM samples in the second OFDM symbol period, then such a path does not exist in the remaining OFDM symbol period. Based on the row-sparsity property, $\boldsymbol{H}^{(2:Q)}_m$'s can be estimated based on the group LASSO technique by solving the following problem
\begin{align} \hspace{-8pt} \mathop{\mathrm{min}}_{\boldsymbol{H}^{(2:Q)}_m}  0.5\left\|\!\boldsymbol{\bar{Y}}_m^{(2:Q)}\!-\!\sqrt{p_m}\boldsymbol{E}_m\boldsymbol{H}^{(2:Q)}_m\!\right\|_{{\rm F}}^2\!+\!\beta\!\sum\limits_{l\in \Phi_m}\!\left\|\boldsymbol{g}_{m,l}\right\|_2, \label{eqn:problem 2}
\end{align}where $\beta\geq 0$ is given. Let $\bar{\boldsymbol{H}}^{(2:Q)}_m=[\bar{\boldsymbol{g}}_{m,1},\ldots,\bar{\boldsymbol{g}}_{m,L}]^T$ denote the optimal solution to the above convex problem, $m=1,2$. Similar to Phase I, given some threshold $\delta_2$, define $\Omega_m^{{\rm II}}=\{l|\|\bar{\boldsymbol{g}}_{m,l}\|_2\geq  \delta_2\}$, $m=1,2$. Then, if $l\in \Omega_m^{{\rm II}}$, we declare that $h_{m,l}^{(q)}\neq 0$, $\forall q\geq 2$. According to (\ref{eq:channel}), each $h_{m,l}^{(q)}$ is contributed by either the BS-target-BS link, or the BS-IRS-BS link, or the BS-IRS-target-BS link. Define $l_m^{{\rm AIA}}$ as the delay (in terms of OFDM samples) corresponding to the BS $m$ to IRS to BS $m$ link that satisfies (\ref{eq:dis 3}), $m=1,2$, and $\Phi_m=\Phi_m^{{\rm I}}\bigcup \{l_m^{{\rm AIA}}\}$, $m=1,2$. As a result, for each $l\in \Omega_m^{{\rm II}}/\Omega_m$, we declare that $h^{{\rm AITA}}_{m,i,1}\neq 0$, $\forall i$, and there exists a target $\hat{k}_{m,l}$ satisfying (\ref{eq:dis 2}). In this case, we estimate $d_{m,\hat{k}_{m,l}}^{{\rm AITA}}$ as the middle point of the range shown in (\ref{eq:dis 2}), i.e.,
\begin{align}\label{eq:range 2}
\bar{d}_{m,\hat{k}_{m,l}}^{\rm AITA}=\frac{(l-1)c_0}{2N\Delta f}+\frac{c_0}{4N\Delta f}, ~ {\rm if} ~ l\in \Phi_m^{{\rm II}}/\Phi_m.
\end{align}To summarize, after the ``on'' state of Phase I, each BS $m$ will have another distance set
\begin{align}\label{eq:set 2}
\mathcal{D}_m^{{\rm AITA}}=\{\bar{d}_{m,\hat{k}_{m,l}}^{\rm AITA}|\forall l\in \Phi_m^{{\rm II}}/\Phi_m\}, ~~~ m=1,2.
\end{align}As a result, after Phase I, the network has four range sets of the targets, i.e., $\mathcal{D}_1^{{\rm AT}}$, $\mathcal{D}_2^{{\rm AT}}$, $\mathcal{D}_1^{{\rm AITA}}$, and $\mathcal{D}_2^{{\rm AITA}}$.

\section{Phase II: Data Association and Localization}\label{sec:Phase II}
In Phase II, we need to localize the $K$ targets based on the knowledge about $\mathcal{D}_1^{{\rm AT}}$, $\mathcal{D}_2^{{\rm AT}}$, $\mathcal{D}_1^{{\rm AITA}}$, and $\mathcal{D}_2^{{\rm AITA}}$ obtained in Phase I. As discussed in Section \ref{sec:Two-Phase Localization Protocol}, the main challenge for localization lies in data association, i.e., we do not directly know how to match each element in the above sets to the right target. For convenience, define $\lambda_{m,k}\in \{1,\ldots,K\}$ such that the estimated distance between BS $m$ and target $k$, i.e., $\bar{d}_{m,k}^{{\rm AT}}$ shown in (\ref{eq:range 1}), is the $\lambda_{m,k}$-th largest element in $\mathcal{D}_m^{{\rm AT}}$, $\forall m,k$. Moreover, define $\mu_{m,k}\in \{1,\ldots,K\}$ such that the estimated distance of the path from BS $m$ to IRS to target $k$ to BS $m$, i.e., $\bar{d}_{m,k}^{{\rm AITA}}$ shown in (\ref{eq:range 2}), is the $\mu_{m,k}$-th largest element in $\mathcal{D}_m^{{\rm AITA}}$, $\forall m,k$. In other words, we have $\bar{d}_{m,k}^{{\rm AT}}=\mathcal{D}_m^{{\rm AT}}(\lambda_{m,k})$ and $\bar{d}_{m,k}^{{\rm AITA}}=\mathcal{D}_m^{{\rm AITA}}(\mu_{m,k})$, $\forall m, k$, where given any set $\mathcal{A}$, $\mathcal{A}(a)$ denotes the $a$-th largest element in $\mathcal{A}$. Note that a feasible data association solution should satisfy
\begin{align}
& \{\lambda_{m,1},\ldots,\lambda_{m,K}\}=\{1,\ldots,K\}, ~ m=1,2, \label{eq:data 1} \\
& \{\mu_{m,1},\ldots,\mu_{m,K}\}=\{1,\ldots,K\}, ~ m=1,2. \label{eq:data 2}
\end{align}

Given the data association, the distance between the IRS and target $k$ estimated by the link from BS $m$ to the IRS to target $k$ to BS $m$ can be expressed as
\begin{align}\label{eq:estimated range IT}
\bar{d}_k^{{\rm IT},m}&=\bar{d}_{m,k}^{{\rm AITA}}-\bar{d}_{m,k}^{{\rm AT}}-d_m^{{\rm AI}} \nonumber \\ &=\mathcal{D}_m^{{\rm AITA}}(\mu_{m,k})\!-\!\mathcal{D}_m^{{\rm AT}}(\lambda_{m,k})\!-\!d_m^{{\rm AI}}, ~ m=1,2.
\end{align}Thus, we have two estimations of $d_k^{{\rm IT}}$, i.e., $\bar{d}_k^{{\rm IT},1}$ via BS 1 and $\bar{d}_k^{{\rm IT},2}$ via BS 2, which should be close to each other.

Define $\mathcal{X}_1=\{\lambda_{m,k},\mu_{m,k}|\forall m, k\}$ and $\mathcal{X}_2=\{(x_k^{{\rm T}},y_k^{{\rm T}})|\forall m,k\}$ as the set of the data association solution and the set of the location solution, respectively. With the estimated distances from each target to BS $1$, to BS $2$, and to the IRS, $\mathcal{X}_1$ and $\mathcal{X}_2$ should be jointly estimated based on the following relation
\begin{align}
& \mathcal{D}_m^{{\rm AT}}(\lambda_{m,k})=\sqrt{(x_k^{\text{T}}-x_m^{\text{A}})^2+(y_k^{\text{T}}-y_m^{\text{A}})^2}+\epsilon_{m,k}, \label{eq:range equation 1} \\
& \mathcal{D}_m^{{\rm AITA}}(\mu_{m,k})-\mathcal{D}_m^{{\rm AT}}(\lambda_{m,k})-d_m^{{\rm AI}}\nonumber \\ &=\sqrt{(x_k^{\text{T}}-x^{\text{I}})^2+(y_k^{\text{T}}-y^{\text{I}})^2}+\varsigma_{m,k}, ~ \forall m,k, \label{eq:range equation 2} \\
& (\ref{eq:data 1}), ~ (\ref{eq:data 2}), \nonumber
\end{align}where $\epsilon_{m,k}$ denotes the error for estimating $d_{m,k}^{{\rm AT}}$, and $\varsigma_{m,k}$ denotes the error for estimating $d_k^{{\rm IT}}$ via the link from BS $m$ to the IRS to target $k$ to BS $m$. In the literature of localization, it is usually assumed that $\epsilon_{m,k}\in \mathcal{CN}(0,\hat{\sigma}_{m,k}^2)$ and $\varsigma_{m,k}\in \mathcal{CN}(0,\tilde{\sigma}_{m,k}^2)$, $\forall m,k$ \cite{Torrieri84,Mao07}.

One straightforward method to estimate $\mathcal{X}_1$ and $\mathcal{X}_2$ is to perform exhaustive search over all the feasible data association solutions that satisfy (\ref{eq:data 1}) and (\ref{eq:data 2}). In particular, given any feasible data association solution $\mathcal{X}_1$, the location of each target $k$ can be obtained by solving the following maximum likelihood (ML) problem \cite{Torrieri84,Mao07}
\begin{align} \! \mathop{\mathrm{min}}_{x_k^{{\rm T}},y_k^{{\rm T}}} ~ \sum\limits_{m=1}^2 (\hat{f}_{m,k}(\lambda_{m,k},x_k^{{\rm T}},y_k^{{\rm T}})+\tilde{f}_{m,k}(\lambda_{m,k},\bar{\mu}_{m,k},x_k^{{\rm T}},y_k^{{\rm T}})), \label{eq:ML problem}
\end{align}where
\begin{align}
& \hat{f}_{m,k}(\lambda_{m,k},x_k^{{\rm T}},y_k^{{\rm T}})\nonumber \\=&\frac{(\mathcal{D}_m^{{\rm AT}}(\lambda_{m,k})-\sqrt{(x_k^{\text{T}}-x_m^{\text{A}})^2+(y_k^{\text{T}}-y_m^{\text{A}})^2})^2}{\hat{\sigma}_{m,k}^2}, \label{eq:cost 1} \\
& \tilde{f}_{m,k}(\lambda_{m,k},\mu_{m,k},x_k^{{\rm T}},y_k^{{\rm T}})\nonumber \\ =&\frac{\!(\!\mathcal{D}_m^{{\rm AITA}}\!(\!\mu_{m,k}\!)\!-\!\mathcal{D}_m^{{\rm AT}}\!(\!\lambda_{m,k}\!)\!-\!d_m^{{\rm AI}}\!-\!\sqrt{\!(\!x_k^{\text{T}}\!-\!x^{\text{I}}\!)\!^2\!+\!(\!y_k^{\text{T}}\!-\!y^{\text{I}}\!)\!^2})^2}{\tilde{\sigma}_{m,k}^2}. \label{eq:cost 2}
\end{align}
Similar to \cite{Torrieri84,Mao07}, we can apply the Gauss-Newton algorithm to solve the above non-convex problem. Let $(x_k^{{\rm T},\ast},y_k^{{\rm T},\ast})$ denote the obtained location of target $k$ corresponding to the given feasible data association solution $\mathcal{X}_1$. Then, the optimal data association solution can be obtained via exhaustive search by solving the following problem
\begin{align}  \mathop{\mathrm{min}}_{\mathcal{X}_1} ~ &  \sum\limits_{k=1}^K\sum\limits_{m=1}^2 (\!\hat{f}_{m,k}(\lambda_{m,k},x_k^{{\rm T},\ast},y_k^{{\rm T},\ast})\nonumber \\ & ~~~~~~~~~~~~+\tilde{f}_{m,k}(\lambda_{m,k},\mu_{m,k},x_k^{{\rm T},\ast},y_k^{{\rm T},\ast})\!) \label{eq:data association problem} \\
\mathrm{s.t.} ~ &  (\ref{eq:data 1}), ~ (\ref{eq:data 2}). \nonumber
\end{align}Given the optimal data association solution, the solution to problem (\ref{eq:ML problem}) can be used as the final location estimation.

However, there are two issues about the performance and complexity of this approach. First, problem (\ref{eq:ML problem}) is a non-convex problem, and it is likely that the Gauss-Newton algorithm will result in a sub-optimal solution. Given the optimal data association, if we obtain a poor localization solution to the non-convex problem (\ref{eq:ML problem}), then this data association solution maybe is not the optimal solution to problem (\ref{eq:data association problem}). In this case, we will obtain a wrong data association solution and the corresponding localization solution is also wrong. Second, the number of feasible data association solutions that satisfy (\ref{eq:data 1}) and (\ref{eq:data 2}) is large, and it is of prohibitive complexity to solve the complicated problem (\ref{eq:ML problem}) given any feasible data association solution, as required by problem (\ref{eq:data association problem}). In the following, we provide an approach that can greatly reduce the number of times to solve the non-convex and complicated problem (\ref{eq:ML problem}).

Specifically, with the knowledge about $\mathcal{D}_1^{{\rm AT}}$, $\mathcal{D}_2^{{\rm AT}}$, $\mathcal{D}_1^{{\rm AITA}}$, and $\mathcal{D}_2^{{\rm AITA}}$, there are two ways to calculate the distance between the IRS and any target $k$: based on the link from BS $1$ to the IRS to target $k$ to BS $1$, i.e., setting $m=1$ in (\ref{eq:estimated range IT}), and based on the link from BS $2$ to the IRS to target $k$ to BS $2$, i.e., setting $m=2$ in (\ref{eq:estimated range IT}). It is worth noting that these two estimations, i.e., $\bar{d}_k^{{\rm IT},1}$ and $\bar{d}_k^{{\rm IT},2}$, should be very close to each other, $\forall k$. As a result, the true data association solution should satisfy
\begin{align}\label{eq:data 3}
& |\mathcal{D}_1^{{\rm AITA}}(\mu_{1,k})-\mathcal{D}_1^{{\rm AT}}(\lambda_{1,k})-\mathcal{D}_2^{{\rm AITA}}(\mu_{2,k})+\mathcal{D}_2^{{\rm AT}}(\lambda_{2,k})|\nonumber \\ \leq & \tau_k, ~ \forall k,
\end{align}where $\tau_k>0$'s are some given thresholds.

Next, define $\mathcal{Y}=\{\mathcal{X}_1|(\ref{eq:data 1}), ~ (\ref{eq:data 2}), ~ (\ref{eq:data 3}) ~ {\rm hold}\}$ as the set of feasible data association solutions. Note that $\mathcal{Y}$ can be obtained with low complexity, because given any data association solution, if there exists a $k$ such that (\ref{eq:data 3}) dose not hold, then this solution is not in the set $\mathcal{Y}$. As a result, we can jointly estimate the data association solution and the location solution as follows. First, given any $\mathcal{D}_1^{{\rm AT}}$, $\mathcal{D}_2^{{\rm AT}}$, $\mathcal{D}_1^{{\rm AITA}}$, and $\mathcal{D}_2^{{\rm AITA}}$, we obtain the set $\mathcal{Y}$. Then, given any data association solution $\mathcal{X}_1\in \mathcal{Y}$, we use the Gauss-Newton algorithm to solve problem (\ref{eq:ML problem}) for obtaining the corresponding location solution $(x_k^{{\rm T},\ast},y_k^{{\rm T},\ast})$'s. Last, the optimal data association solution is obtained by solving the following problem
\begin{align}  \mathop{\mathrm{min}}_{\mathcal{X}_1} ~ &  \sum\limits_{k=1}^K\sum\limits_{m=1}^2 (\!\hat{f}_{m,k}(\lambda_{m,k},x_k^{{\rm T},\ast},y_k^{{\rm T},\ast})\nonumber \\ & ~~~~~~~~~~~~+\tilde{f}_{m,k}(\lambda_{m,k},\mu_{m,k},x_k^{{\rm T},\ast},y_k^{{\rm T},\ast})\!) \label{eq:data association problem 1} \\
\mathrm{s.t.} ~ &  (\ref{eq:data 1}), ~ (\ref{eq:data 2}), ~ (\ref{eq:data 3}). \nonumber
\end{align}Given the optimal data association solution, the solution to problem (\ref{eq:ML problem}) can be used as the final estimation of the location of each target. Because not too many data association solutions can satisfy (\ref{eq:data 3}) in practice, we only need to perform exhaustive search over a very small set in problem (\ref{eq:data association problem 1}). As a result, the complexity of the proposed algorithm is very low. Moreover, since data association under the proposed scheme relies more on the simple criterion (\ref{eq:data 3}), rather than the solution to the non-convex problem (\ref{eq:ML problem}) that may be sub-optimal, the data association and localization accuracy is expected to be improved significantly.

\section{Discussion: 2 BSs Plus 1 IRS v.s. 3 BSs}
Recently, \cite{dvc} proposed a device-free sensing architecture based on the trilateration technique, where all the anchors are active BSs. Interestingly, the results in this paper show that two active BSs and one more passive IRS are sufficient to enable device-free sensing based on the trilateration technique. There are several advantages to deploy two BSs and one IRS for localization as compare to deploying three BSs. First, the IRS is of lower cost and can be deployed at more sites compared to the BS. Second, the distance between an BS and a target can be merely measured by estimating the delay of the BS-target path, while the distance between the IRS and a target can be measured by two ways as shown in (\ref{eq:estimated range IT}), because the signals reflected by the IRS can be received by both BSs at their assigned sub-carriers, as shown in (\ref{eq:rcvd_sgnl}). This redundant information enables a low-complexity and high-quality data association algorithm that does not significantly rely on the solution to the non-convex problem (\ref{eq:ML problem}). However, there are also limitations for the IRS-enabled localization. For example, this architecture can only localize the targets that are close to the IRS such that their reflected signals via the IRS is strong enough to be detected by the BSs.

\vspace{-5pt}
\section{Numerical Results}\label{sec:Numerical Results}
In this section, we provide numerical results to verify the effectiveness of the proposed two-phase localization protocol. It is assumed that BS $1$ and BS $2$ are localized at $(-100,0)$ in meter and $(100,0)$ in meter, respectively, and the IRS is located at $(0,40)$ in meter. All the targets are randomly localized within a circle whose center is the IRS site and radius is $50$ meters. Moreover, the channel bandwidth is $400$ MHz, and the number of OFDM symbols in a RB is $Q=7$. Last, the identical transmit power of the BSs is $39$ dBm, and the power spectrum density of the noise at the BSs is $-174$ dBm/Hz.

\begin{figure}[t]
  \centering
  \includegraphics[width=8cm]{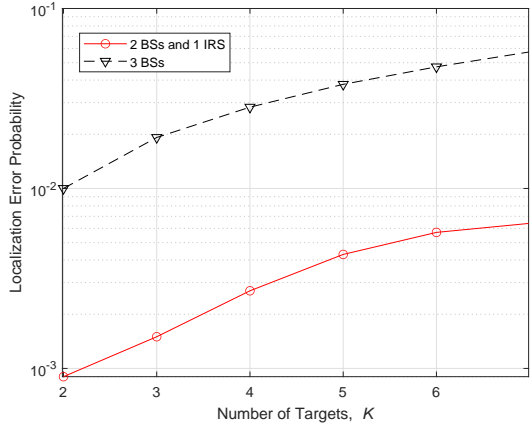}\vspace{-5pt}
  \caption{Localization performance comparison: the setup with 2 BSs and 1 IRS considered in this paper versus the setup with 3 BSs considered in \cite{dvc}.}\label{fig2}\vspace{-15pt}
\end{figure}

Fig. \ref{fig2} shows the localization error probability achieved by our proposed two-phase protocol in the considered network with 2 BSs and 1 IRS. Here, an error event for localizing a target is defined as the case that the estimated location is not lying within a radius of $1$ meter from the true target location. For performance comparison, we consider the cellular network consisting of 3 BSs as the benchmark, where the localization can be performed by utilizing the scheme proposed in \cite{dvc}. It is observed that because the data association is based on the simple criterion (\ref{eq:data 3}) rather than the solution to the non-convex problem (\ref{eq:ML problem}) as in \cite{dvc}, our proposed scheme can achieve lower localization error probability over the benchmark scheme. Moreover, numerical results also verify that our proposed algorithm is of lower complexity. For example, when $K=7$, the average CUP running time of the benchmark scheme is about $0.09$ s for each realization, while that of the proposed scheme is about $0.01$ s.

\vspace{-5pt}
	
\section{Conclusions}
In this paper, we considered the trilateration-based device-free sensing in a cellular network consisting of two BSs, one IRS, and multiple passive targets. Compared to the device-based sensing counterpart with three active anchors and multiple active targets, there are two main challenges in the considered network. First, it is difficult to measure the distance between the passive IRS and each passive target because both of them cannot estimate the delay for the signal propagated between them. Second, it is non-trivial to match each estimated range to the right target. Our results showed that the above challenges can be efficiently tackled based on the advanced signal processing techniques. As a result, it is concluded that the trilateration-based technique can be generalized to the case when some of the anchors and all the targets are passive.

\vspace{-5pt}
\bibliographystyle{IEEEtran}
\bibliography{ISAC}
	
\end{document}